\documentstyle[12pt]{article}
\def\bq{\begin{eqnarray}}
\def\eq{\end{eqnarray}}
\def\prt{\partial}

\begin{document}
\title{\bf Dynamical mass shift for a partially reflecting moving mirror}
\author{NISTOR NICOLAEVICI\\
\it Technical University of Timi\c soara, Department of Physics\\
\it P-\c ta. Hora\c tiu 1, RO-1900 Timi\c soara, Romania}                               
\maketitle
\begin{abstract}

We consider the vacuum fluctuations contribution to the mass of a mirror in an 
exactly soluble partially reflecting moving mirror model. Partial reflectivity 
is accounted for by a repulsive delta-type potential localized along the 
mirror's trajectory. The mirror's mass is explicitly obtained as an integral 
functional of the mirror's past trajectory.   
\end{abstract}

\section{Introduction}\label{one}

Some years ago in a series of papers \cite{jaekel1}-\cite{jaekel5} Jaekel and 
Reynauds discussed quantum field theory implications to the inertial properties 
of partially reflecting mirrors. They evidenced that vacuum fluctuations lead 
to a shift in the mirror's mass, which is not happening for perfectly 
reflecting mirrors. The phenomenon can be intuitively understood by recalling 
a well-known situation in non-relativistic wave mechanics: scattering on barrier 
potentials is generally accompanied by a time delay, implying that for a certain 
time the energy of the scattered object can be considered to contribute to the 
proper energy of the scatterer. The mirror (taken here as a classical object) in 
the vacuum of the quantized field can be viewed as scattering the 
virtual zero-point oscillations. For perfect reflectors there is no time delay 
and thus no mass shift. 

A particularly simple case evidencing the apparition of the mass shift is that 
of a pointlike mirror in 1+1 Minkowski space with the mirror-field 
interaction strictly localized at the mirror's position \cite{jaekel4, jaekel5}. 
The model was discussed in the cited papers basically at the 
classical level. Our intention here is to analyze the quantized version. 
We shall base our investigation on the results 
obtained in Ref. \cite{nicola} where quantization of the massless scalar 
field corresponding to this model was achieved. The treatment 
therein has the convenient features (i) of being non-perturbative, and (ii) 
of relying on the explicitly constructed space-time dependent Heisenberg 
field operator. These make possible the $exact$ evaluation of local 
quantities, such as the radiated energy-momentum density or correlation 
functions, at $arbitrary$ times (by contrast, in Refs. 
\cite {jaekel1}-\cite{jaekel5} an in-out formalism\footnote{See Refs. 
\cite{jaekel5bis}-\cite{jaekel7}.} was used, and calculations were mainly 
performed in the small mirror displacements approximation).

Our basic result is the non-perturbative trajectory dependence of the 
mass shift. We take the view that the free motion quantity is ultimately 
unobservable, similar to the electromagnetic mass shift for the electron in 
Q.E.D., say. Hence, by definition uniform trajectories will be characterized 
by vanishing mass shifts. Practically, the model implies that the mass shift is 
given by the renormalized two point function evaluated at coincidence points at 
the mirror's instantaneous position. One naturally interprets it as originating 
in the distortion of the vacuum fluctuations produced by the mirror's 
(non-uniform) motion. For perfectly reflecting mirrors, absence of the mass 
shift can be equivalently seen as due to the ``freezing'' of the vacuum 
fluctuations along the trajectory, as the field operator itself vanishes at 
these points \cite{fulling, davies}. A most notable feature is that the mass 
shift appears as a history dependent quantity. This is directly related to the 
same characteristic displayed by the quantum flux radiated by the mirror 
\cite{nicola}.

Another issue we discuss is the connection with the mirror dynamical 
equation, taking into account the quantum field backreaction. We point 
out that the mass shift appears from the lack of orthogonality between the 
mirror two-velocity and the backreaction 
force. Perfect mirrors prove to have this relation assured \cite{fulling}.

The paper is organized as follows. In Sec. \ref{two} we briefly describe the 
mirror model. In Sec. \ref{three} we relate the mass shift to the 
mirror dynamics. Considerations in this section are rather general, as they 
do not refer to a particular mirror model. In Sec. \ref{four} we 
obtain the trajectory dependence and evidence some of its 
general properties. Section \ref{five} contains discussions concerning the 
positivity of the mass shift, which is left as an open problem.

\section{The mirror model}\label{two}
Let $z^\mu$ $(\mu=0,1)$ represent the coordinates in the $1+1$ Minkowski 
space\footnote{The metric tensor is $g_{00}=-g_{11}=1$. Natural units 
$\hbar=c=1$ are used throughout the paper.} and $z^\mu(\tau)$ denote the mirror 
trajectory with $\tau$ the proper time 
\bq
d\tau=\sqrt{d z^\mu dz_{\mu}}.
\eq
The mirror-field system is described by the action
\bq
S[\phi,z]=\frac{1}{2}\int d^2z\prt_\mu \phi(z)\prt^\mu\phi(z)
-m\int d\tau-\frac{a}{2}\int d\tau\phi^2(z(\tau)),\label{action}
\eq
where $m$ is the mirror's mass and $a$ is a positive constant characterizing the 
mirror-field interaction. The field equation following from (\ref{action}) is
\bq
(\Box+V(z))\phi(z)=0,\label{fe}
\eq
with 
\bq
V(z^0,z^1)=\frac{a}{\dot z^0(\tau)}\delta (z^1-z^1(\tau_z)),
\label{vpot}
\eq
where $\delta$ denotes the Dirac distribution and $\tau_z$ is implicitely 
determined from equation $z^0(\tau_z)=z^0$. The overdot represents 
differentiation with respect to proper time. Eq. (\ref{vpot}) is equivalent 
to a repulsive barrier potential localized on the trajectory with $a$ 
correponding to the barrier strength. One obtains thus a semitransparent moving mirror 
model with semitransparency controlled by parameter $a$.

Quantization of $\phi (z)$ respecting eq. (\ref{fe}) for 
$z^\mu(\tau)$ arbitrary was performed in Ref. \cite{nicola}. 
We refer to this paper for the expressions of the in-vacuum two point function 
and the renormalized energy-momentum tensor.

Now, translational invariance of $S$ assures conservation of a total mirror-field 
energy-momentum
\bq
P_\mu=\int_{-\infty}^{+\infty}dz^1 \,T_{\mu 0}
(z)+
[\,m+\frac{a}{2}\phi^2(z(\tau))\,]\,\dot z_\mu(\tau),
\label{total}
\eq
where
\bq
T_{\mu\nu}=\prt_\mu\phi\prt_\nu\phi-g_{\mu\nu}\prt^\alpha\phi\prt_\alpha\phi,
\label{tensor}
\eq
and $\tau$ is arbitrary. Integration is taken along the spatial hypersurface 
$z^0=z^0(\tau)$. One sees that the interaction term is 
equivalent to a $\phi$ dependent contribution to the mirror's mass  
\bq
\mu=\frac{a}{2}\phi^2.
\label{shiftop}
\eq
With $\phi$ the quantum field, eq. (\ref{shiftop}) defines the mass 
shift operator, as already discussed in Refs. \cite{jaekel4, jaekel5}.

We shall assume in the following that the quantum field is in the 
vacuum state at infinite past. This ammounts to replace all 
operatorial quantities with their corresponding renormalized in-vacuum expectation values. 
We shall understand this tacitly in what follows, without a notational change. 
Our interest lies in the trajectory dependence of $\mu(\tau)$. Before 
going into details, we consider next the connection with the 
dynamical equation.

\section {Dynamical equation and mass shift}\label{three}

Consider for the moment the free field contribution in eq. (\ref{total}). 
We rewrite it as
\bq
P_\mu^{field}(\tau)=\int_{-\infty}^{z^1_-(\tau)}dz^1\,T^L_{\mu0}(z)+
\int_{z^1_+(\tau)}^{+\infty} dz^1\,T^R_{\mu0}(z), 
\label{field}
\eq
where 
\bq
z^1_\pm(\tau)=z^1(\tau)\pm \epsilon,
\label{ep}
\eq
with $\epsilon >0$, $\epsilon \rightarrow 0$. Superscripts $L$, $R$ 
refer to the left and right regions of the Minkowski 
plane, as naturally determined by the trajectory. We excluded from 
integration an infinitesimal vecinity around $z^1(\tau)$, with no 
relevance for the total quantity. Note that the barrier potential implies that 
for points strictly on the trajectory $\prt_\mu\phi$ are discontinous, while 
$T_{\mu\nu}$ has a non-vanishing divergence.

The $\tau$-derivative of eq. (\ref{field}) is minus the two-force acting 
on the mirror due to the field backreaction. In null coordinates 
$z^{\pm}=z^0\pm z^1$ one has
\bq
\dot P_+^{field}(\tau)=T_{++}^L(z(\tau))\, \dot z^+(\tau),\label{v}\\
\dot P_-^{field}(\tau)=T_{--}^R(z(\tau))\, \dot z^-(\tau),\label{u}
\eq
where $z(\tau)$ in $T^L_{++}$ is understood as 
\bq
(z^0(\tau),z^1_-(\tau)),
\eq
and analogously with $z^1_-\rightarrow z^1_+$ for $T^R_{--}$. 
To obtain relations above one uses first the divergenceless of 
$T_{\mu\nu}$ $inside$ the $L$, $R$ regions to reduce the $z_1$-integrations 
to pure boundary terms at $z(\tau)$. One further takes into account that some 
of the $T_{\mu\nu}$ components vanish. By eq. (\ref{tensor}) one has
\bq
T^{R,L}_{+-}=T^{R,L}_{-+}=0.
\label{uv}
\eq
Explicit calculations also show
\bq
T_{++}^R=T_{--}^L=0.
\label{**}
\label{noflux}
\eq
(This states that there is no incoming flux from past null infinity, a direct 
consequence of choosing the field in the vacuum state at infinite past.) Using 
now total energy-momentum conservation one obtains
\bq
\frac{d}{d\tau}[m(\tau)\dot z_+(\tau)]+
T_{++}^L(z(\tau))\, \dot z^+(\tau)=0\label{veq}\\
\frac{d}{d\tau}[m(\tau)\dot z_-(\tau)]+
T_{--}^R(z(\tau))\, \dot z^-(\tau)=0\label{ueq},
\eq
were we introduced the total mass
\bq
m(\tau)=m+\mu(\tau).
\label{totalmass}
\eq
One can further eliminate $\ddot z^\pm$ from eqs. (\ref{veq}), 
(\ref{ueq}) by using the othogonality relation 
\bq
\ddot z_+\dot z^++\ddot z_-\dot z^-=0. 
\eq
One finds 
\bq
\dot \mu(\tau)+T_{++}^L(z(\tau))\,\dot z^+(\tau)^2 +
T_{--}^R(z(\tau))\,\dot z^-(\tau)^2=0.\label{mu}
\eq
Hence, $\mu$ variations compensate for the non-orthogonality between the mirror 
velocity and the backreaction force.

Quantities $T^R_{--}$, $T^L_{++}$ were explicitly obtained in 
Ref. \cite{nicola} as integral functionals of the mirror's past trajectory. When 
evaluated at $z(\tau)$ 
as above, trajectory dependence extends up to the proper time $\tau$, in agreement 
to causality. Eqs. (\ref{veq})-(\ref{totalmass}) and (\ref{mu}) provide an 
integro-differential 
system determining the mirror dynamics (in the absence external forces). For our 
discussion, it is relevant eq. (\ref{mu}). It defines the mass shift in 
terms of the past mirror's trajectory, independently of the dynamical problem. 
We point out that eq. (\ref{mu}) could have been equivalently obtained from 
only manipulating the field equation (\ref{fe}) and using eq. (\ref{**}). 
Considerations above were merely intended to make clear the dynamical origin of 
$\mu$.

\section{Trajectory dependence of $\mu(\tau)$}\label{four}

Eq. (\ref{shiftop}) defines the mass shift in terms of the renormalized two point 
function evaluated at identical points $z(\tau)$. This yields\footnote{See 
Ref. \cite{nicola}, eqs. (36)-(39). Please note that quantities therein are 
written for $a\rightarrow 2a$. $i\epsilon$ prescription is unnecessary here.} 
\bq
 \mu=\mu_1+\mu_2,\label{wig1}
\eq
where 
\bq
\nonumber\\
 \mu_1(\tau)=\frac{a}{8\pi}
      \int_{-\infty}^{\tau}d\tau_1
 \lbrace\,
 \ln ((z^-(\tau_1)-z^-(\tau))(z^+(\tau_1)-z^+(\tau))\,
\nonumber\\
\nonumber\\
 \times\exp\,a(\tau_1-\tau)/2\, \rbrace
\hspace*{39pt}
\nonumber\\
\nonumber\\
 +\frac{a}{8\pi}\int_{-\infty}^{\tau}d\tau_1
 \lbrace\,
 \ln ((z^-(\tau)-z^-(\tau_1))(z^+(\tau)-z^+(\tau_1))
\nonumber\\
\nonumber\\
 \times\exp\,a(\tau_1-\tau)/2\,\rbrace,
\hspace*{30pt}\label{wig2}
\end{eqnarray}
\begin{eqnarray}
 \mu_2(\tau)= -\frac{a^2}{16\pi}
     \int_{-\infty}^{\tau}
     \int_{-\infty}^{\tau}
         d\tau_1 d\tau_2\lbrace\,\ln 
 ((z^+(\tau_1)-z^+(\tau_2))\hspace*{8pt}
\nonumber\\
\nonumber\\
 \times (z^-(\tau_1)-z^-(\tau_2))\,
 \exp\,a((\tau_1+\tau_2)/2-\tau)\,\rbrace.
\label{wig3}
\nonumber\\
\end{eqnarray}
Past history dependence is manifest. On the other side, the $\tau$-derivative 
of $\mu$ is determined by eq. (\ref{mu}) 
in terms of the renormalized energy-momentum tensor. One 
finds\footnote{$ibidem$, eqs. (42)-(45).} 
\bq
\dot \mu(\tau)=-\frac{a}{8\pi}\int_{-\infty}^{\tau}
\int_{-\infty}^{\tau} 
d\tau_1d\tau_2 \prt_{\tau_1}\prt_{\tau_2}
\left(
\frac{\dot z^+(\tau_1)-\dot z^+(\tau_2)}
{z^+(\tau_1)-z^+(\tau_2)}
+
\frac{\dot z^-(\tau_1)-\dot z^-(\tau_2)}
{z^-(\tau_1)-z^-(\tau_2)}
\right)\nonumber\\
\times\exp\, a((\tau_1+\tau_2)/2-\tau).
\hspace*{20pt}
\label{muderiv}
\eq
A straightforward calculation shows that eq. (\ref{muderiv}) results indeed 
from eqs. (\ref{wig1})-(\ref{wig2}). We want to stress out that this does not 
trivially follows from eqs. (\ref{tensor}), (\ref{shiftop}), (\ref{mu}), 
seen as defining an operatorial identity for $\phi$. Quantities above result 
from a renormalization procedure, implying the potentially dangerous infinite 
subtraction.

We evidence next some general properties of $\mu(\tau)$ following from 
eqs. (\ref{wig1})-(\ref{muderiv}).

One sees from eq. (\ref{muderiv}) that $\mu$ is constant 
along uniform trajectories. Explicit calculations using eqs. 
(\ref{wig1})-(\ref{wig3}) yield the velocity-independent quantity
\bq
\mu_0=\frac{a}{4\pi}\left(-\ln \frac{a}{2}+
\int_{0}^{\infty}dx\ln x\, e^{-x}
\right).\label{expl}
\eq
As mentioned in Sec. \ref{one}, we regard $\mu_0$ as an inobservable 
contribution to the physical mirror mass. Correspondingly, we redefine 
the mass shift as
\bq
\mu(\tau)\rightarrow \mu(\tau)-\mu_0.
\label{redef}
\eq
Note that by eqs. (\ref{wig2}), (\ref{wig3}) $\mu$ contains the 
logarithm of a dimensionful quantity. This is remediated 
by subtraction (\ref{redef}). 
We recall that in quantizing $\phi$ the trajectories were supposed 
uniform in the infinite past. This is physically not 
a serious restriction and can be always assumed in a realistic situation. 
Combined with eq. (\ref{redef}), it is clear that this requests
\bq
\lim_{\tau\rightarrow -\infty}\mu(\tau)=0,\label{cond}
\eq
which serves as the initial condition for eq. (\ref{muderiv}).

The exponentials in $\mu_1$, $\mu_2$ imply that $\mu(\tau)$ is 
essentially determined by the motion before $\tau$ in an interval of 
order $\sim a^{-1}$. This has an immediate consequence. Consider a 
trajectory with the velocity remaining constant after a fixed 
proper time $\tau_0$. One has then that for $\tau-\tau_0 
\gg a^{-1}$ the mass shift practically vanishes, as in the integrals (\ref{wig2}), 
(\ref{wig3}) the trajectory can be approximated with an uniform one. 

The perfect reflectivity limit is obtained\footnote{$ibidem$.} by making 
$a\rightarrow \infty$. It's not hard to see that eq. (\ref{muderiv}) implies 
$\lim_{a\rightarrow \infty}\dot \mu(\tau)=0$. Taking into account condition 
(\ref{cond}) one concludes that the mass shift is absent for perfect mirrors, as 
pointed out in Sec. \ref{one}.

It is relevant to consider the case of slowly varying motions on a proper time 
scale of order $a^{-1}$, i.e.
\bq
\vert \alpha  \vert \ll a^{-1},\quad 
\biggl \vert \frac{d^{n+1}\alpha}{d\tau^{n+1}}\biggl /
\frac{d^{n}\alpha}{d\tau^{n}}\biggl \vert \ll a^{-1},\quad 
n=0,1,2,\dots,\label{ineq}
\eq
where $\alpha$ denotes the mirror's proper acceleration. This is equivalent to 
``large'' values for $a$, corresponding to a near perfect mirror behaviour. Then 
the paranthesis in eq. (\ref{muderiv}) is also a slowly varying function on 
intervals $\Delta \tau_{1,2}\sim a^{-1}$. Assuming the trajectory 
infinitely differentiable, one may Taylor expand it around $\tau_1=\tau_2=\tau$ 
and perform a term by term integration. The result is a $a^{-n}$ ($n\geq 1$) 
expansion 
with the coefficients entirely 
determined by the acceleration and its derivatives evaluated at $\tau$. We 
give below the $O(a^{-3})$ contribution 
\bq
\dot \mu(\tau)=\frac{1}{24\pi}\left [
\,\frac{1}{a}\alpha \dot \alpha-
\frac{1}{a^2}
(\alpha \ddot \alpha+\dot \alpha^2)+\dots\, \right ]
_{\tau}.\label{muap}
\eq
One remarks that the expression above can be integrated to yield $\mu$ 
itself as a purely local quantity in terms of $\alpha$, $\dot \alpha$. 
Restricting to the leading order, one has the interesting result
\bq
\mu(\tau)=\frac{
\alpha^2(\tau)}
{48\pi a}
.\label{muacc}
\eq

Eq. (\ref{muap}) suggests that $\dot\mu$ is zero on uniformly 
accelerated trajectories $\alpha=$const. One can write in this case 
\bq
z^{+}(\tau_1)-z^{+}(\tau_2)=A\exp B(\tau_1-\tau_2),
\label{acc1}\\
z^{-}(\tau_1)-z^{-}(\tau_2)=\frac{1}{A}\exp \frac{1}{B}(\tau_1-\tau_2),
\label{acc2}
\eq
with $A$, $B$ fixed and of identical sign. 
One easily verifies that for eqs. (\ref{acc1}), (\ref{acc2}) the integrand 
in eq. (\ref{muderiv}) vanishes. This implies that constant accelerations 
maintained on intervals significantly larger than $a^{-1}$ lead 
to a constant mass shift. This is of course not to say that $\mu$ vanishes on 
these trajectories; it just means that variations of 
the mass shift are mainly located in the transient phases between pieces 
of the trajectories with constant acceleration (if the case).

Finally, we mention that the mass shift is a purely relativistic effect: 
in the non-relativistic approximation of the parantheses in eq. 
(\ref{muderiv}), terms linear in velocity turn out to cancel among themselves.

\section{Discussion}\label{five}
We left unanswered the essential question concerning the sign of $\mu(\tau)$. 
For sufficiently slowly varying motions, the leading order contribution 
(\ref{muacc}) is always non-negative. Adding the higher orders is 
not expected to yield 
a negative quantity, due to the inequalities (\ref{ineq}). (Note also that 
eq. (\ref{muap}) says that if the $a^{-1}$ contribution vanishes, so does 
the $a^{-2}$ one). One may be inclined to 
think that positivity of $\mu$ is connected to slowly varying motions. This is 
not necessarilly so. Consider the trajectory with constant velocities $\beta_i$ for 
$\tau<0$ and $\beta_f$ for $\tau>0$. Then for proper times 
$0<a\tau\ll 1$ one finds
\bq
\mu(\tau)=\frac{a}{4\pi}
\left[
\gamma_i \gamma_f(1-\beta_1\beta_2)-1
\right]
(-a\tau \ln a\tau+O(a\tau))>0,
\eq
with $\gamma_{\,i,f}=(1-\beta_{\,i,f}^2)^{-1/2}$. More generally, one can show 
that the inequality above holds for arbitrary non-uniform $\tau>0$ trajectories, 
for $\tau$ sufficiently small.

We haven't succeeded in finding trajectories with $\mu$ assuming negative values. 
We have neither found a proof to forbid this possibility. In Refs. \cite{jaekel1,jaekel6} it was 
claimed that the mass shift is always positive, irrespective of the mirror model (within 
the assumption of causality and unitarity). The conclusion was 
obtained, however, in the context of ignoring renormalization. 
Note that this implies in eq. (\ref{shiftop}) the subtraction of the 
infinite $positive$ free field contribution. In this respect the situation is 
principially the same, e.g., with that of the energy density radiated by a perfectly 
reflecting moving mirror \cite{fulling,davies}. The corresponding operator is formally 
positively 
defined, but after renormalization negative values are allowed for certain 
trajectories. If a similar situation happens here, one may 
contemplate the possibility of generating negative masses, entailing the 
well-known unphysicalities. (For example, one could imagine the $-m$ negative 
mass mirror attached to an object of mass $m$ not interacting with the 
$\phi$ field, both stabilized on an uniformly accelerated trajectory in the 
absence of external forces; this would represent a system converting the vacuum 
zero-point energy into the kinetic energy of the accelerated 
object\footnote{
Theoretically, this is the case for the run-away trajectories of perfectly 
reflecting mirrors \cite{jaekel8,jaekel9}. This is realistically not expected due to 
the unphysical assumption of perfect reflectivity. When partial reflectivity is 
taken in to account run-away trajectories are forbidden, see the cited papers.}.) We leave open the question of the mass shift sign, hoping to return 
with further results. 

\bigskip
\noindent
\section*{Acknowledgments}
I thank to Farkas Atilla for stimulating discussions and for reading the 
manuscript.ÿÿ
ÿÿÿ

\end{document}